\pdfoutput=1
\RequirePackage{ifpdf}
\ifpdf % We are running pdfTeX in pdf mode
\documentclass[pdftex]{sigma}
\else
\documentclass{sigma}
\fi

\begin{document}

\allowdisplaybreaks

\renewcommand{\thefootnote}{$\star$}

\renewcommand{\PaperNumber}{073}

\FirstPageHeading

\ShortArticleName{Application of the BDEs Method to One Problem of Free Turbulence}

\ArticleName{Application of the $\boldsymbol{B}$-Determining Equations Method\\ to One Problem of Free Turbulence\footnote{This
paper is a contribution to the Special Issue ``Geometrical Methods in Mathematical Physics''. The full collection is available at \href{http://www.emis.de/journals/SIGMA/GMMP2012.html}{http://www.emis.de/journals/SIGMA/GMMP2012.html}}}

\Author{Oleg V.~KAPTSOV and Alexey V.~SCHMIDT}

\AuthorNameForHeading{O.V.~Kaptsov and A.V.~Schmidt}

\Address{Institute of Computational Modeling SB RAS, Akademgorodok, Krasnoyarsk, 660036, Russia}
\Email{\href{mailto:kaptsov@icm.krasn.ru}{kaptsov@icm.krasn.ru}, \href{mailto:schmidt@icm.krasn.ru}{schmidt@icm.krasn.ru}}

\ArticleDates{Received May 17, 2012, in f\/inal form October 04, 2012; Published online October 16, 2012}

\Abstract{A three-dimensional model of the far turbulent wake behind a self-propelled body in a passively stratif\/ied medium is considered. The model is reduced to a system of ordinary dif\/ferential equations by a similarity reduction and the $B$-determining equations method. The system of ordinary dif\/ferential equations satisfying natural boundary conditions is solved numerically. The solutions obtained here are in close agreement with experimental data.}


\Classification{76M60; 76F60}

\renewcommand{\thefootnote}{\arabic{footnote}}
\setcounter{footnote}{0}

\section{Introduction}

{\sloppy Most f\/lows occurring in nature and engineering practice are turbulent (see, e.g., \cite{Hinze,Pope,Schlichting}). Semiempirical models of turbulence are widely used in the modeling of turbulent f\/lows \mbox{\cite{LS,Rodi,Wilcox}}. However, there are only a few analytical results on such models (see, e.g.,~\cite{Barenblatt,Hulshof}).

}

The far turbulent wake behind an axisymmetric body in a stratif\/ied medium is an example of a free shear f\/low. Suf\/f\/iciently complete experimental data on the dynamics of turbulent wakes generated by moving bodies in stratif\/ied f\/luids were obtained by Lin and Pao~\cite{LP}.

The far turbulent wake behind an axisymmetric towed body in a linearly stratif\/ied medium was numerically simulated in \cite{Hassid}. Chernykh et al.~\cite{Ch1} carried out the numerical simulation of the dynamics of turbulent wakes in a stable stratif\/ied medium based on hierarchy of second order closure models. Calculation results obtained in~\cite{Ch1,Hassid} are in close agreement with experimental data \cite{LP}.

Similarity solutions for several turbulence models were constructed in \cite{TurbSymm3,TurbSymm1,TurbSymm2,Shanko}. In the current paper we consider three-dimensional semiempirical model  of the far turbulent wake behind an axisymmetrical self-propelled body in a passively stratif\/ied medium (see \cite{Ch1,Ch2,Ch3} and the references therein).

This paper is organized as follows. In Section~\ref{section3} we determine the most general continuous classical symmetry group of the model and obtain the similarity reduction of the model. In Section~\ref{section4} we use the $B$-determining equations (BDEs) method~\cite{BDE,BDE_Paper} to transform the reduced system into a system of ordinary dif\/ferential equations~(ODEs).

In the last section, we consider a boundary value problem for the system of ODEs. We use the modif\/ied shooting method and the asymptotic expansion of the solution in the vicinity of  the singular point to solve this problem. Finally, computational results are given.

\section{Model}\label{section2}

The following three-dimensional semiempirical model of turbulence was constructed in \cite{Ch2,Ch1,Ch3} to calculate characteristics of the far turbulent wake behind an axisymmetric self-propelled body in a passively stratif\/ied medium:
\begin{gather}
U_0\frac{\partial e}{\partial x}=\frac{\partial}{\partial y}C_e\frac{e^2}{\varepsilon}\frac{\partial e}{\partial y}+\frac{\partial}{\partial z}C_e\frac{e^2}{\varepsilon}\frac{\partial e}{\partial z}-\varepsilon,\label{equation1}\\
U_0\frac{\partial \varepsilon}{\partial x}=\frac{\partial}{\partial y}C_\varepsilon \frac{e^2}{\varepsilon}\frac{\partial \varepsilon}{\partial y}+\frac{\partial}{\partial z}C_\varepsilon \frac{e^2}{\varepsilon}\frac{\partial \varepsilon}{\partial z}-{C_\varepsilon}_2\frac{\varepsilon^2}{e},\label{equation2}\\
U_0\frac{\partial \langle\rho_1\rangle}{\partial x}=\frac{\partial}{\partial y}C_\rho\frac{e^2}{\varepsilon}\frac{\partial \langle\rho_1\rangle}{\partial y}+\frac{\partial}{\partial z}C_\rho\frac{e^2}{\varepsilon}\frac{\partial \langle\rho_1\rangle}{\partial z}-\frac{\partial }{\partial z}C_\rho\frac{e^2}{\varepsilon},\label{equation3}\\
U_0\frac{\partial \langle\rho'^2\rangle}{\partial x}=\frac{\partial}{\partial y}{C_1}_\rho\frac{e^2}{\varepsilon}\frac{\partial \langle\rho'^2\rangle}{\partial y}+\frac{\partial}{\partial z}{C_1}_\rho\frac{e^2}{\varepsilon}\frac{\partial \langle\rho'^2\rangle}{\partial z}+2C_\rho\frac{e^2}{\varepsilon}{\frac{\partial \langle\rho_1\rangle}{\partial y}}^2\nonumber\\
\hphantom{U_0\frac{\partial \langle\rho'^2\rangle}{\partial x}=}{}
 +2C_\rho\frac{e^2}{\varepsilon}\left(\frac{\partial \langle\rho_1\rangle}{\partial z}-1\right)^2-C_T\frac{\langle\rho'^2\rangle\varepsilon}{e},\label{equation4}
\end{gather}
where $e$ is the turbulent kinetic energy, $\varepsilon$ is the kinetic energy dissipation rate, $\langle\rho_1\rangle$ is the averaged density defect, and $\langle\rho'^2\rangle$ is the density f\/luctuation variance. All the unknown functions depend on $x$, $y$, and $z$. The quantities $C_e=0.136$, $C_\varepsilon=C_e/\delta$, $\delta=1.3$, ${C_\varepsilon}_2=1.92$, $C_\rho=0.208$, ${C_1}_\rho=0.087$, $C_T=1.25$ are generally accepted empirical constant \cite{GL,Rodi}. $U_0$ is the free stream velocity. The marching variable $x$ in the equations \eqref{equation1}--\eqref{equation4} acts as time.

This model is based on the three-dimensional parabolized system of averaged Navier--Stokes equations in the Oberbeck--Boussinesq approximation (see \cite{Ch_NS1,Ch_NS2})
\begin{gather}
U_0\frac{\partial U_d}{\partial x}+V\frac{\partial U_d}{\partial y}+W\frac{\partial U_d}{\partial z}=\frac{\partial}{\partial y}\langle u'v'\rangle+\frac{\partial}{\partial z}\langle u'w'\rangle,\label{equation5}\\
U_0\frac{\partial V}{\partial x}+V\frac{\partial V}{\partial y}+W\frac{\partial V}{\partial z}=-\frac 1{\rho_0}\frac{\partial \langle p_1 \rangle}{\partial y}-\frac{\partial}{\partial y}\langle v'^2\rangle-\frac{\partial}{\partial z}\langle v'w'\rangle,\label{equation6}\\
U_0\frac{\partial W}{\partial x}+V\frac{\partial W}{\partial y}+W\frac{\partial W}{\partial z}=-\frac 1{\rho_0}\frac{\partial \langle p_1 \rangle}{\partial z}-\frac{\partial}{\partial y}\langle v'w'\rangle-\frac{\partial}{\partial z}\langle w'^2\rangle-g\frac{\langle \rho_1 \rangle}{\rho_0},\label{equation7}\\
U_0\frac{\partial \langle\rho_1\rangle}{\partial x}+V\frac{\partial \langle\rho_1\rangle}{\partial y}+W\frac{\partial \langle\rho_1\rangle}{\partial z}+W\frac{\partial \rho_s}{\partial z}=-\frac{\partial}{\partial y}\langle v'\rho'\rangle-\frac{\partial}{\partial z}\langle w'\rho'\rangle,\label{equation8}\\
\frac{\partial V}{\partial y}+\frac{\partial W}{\partial z}=\frac{\partial U_d}{\partial x},\label{equation9}
\end{gather}
where $U_d=U_0-U$ is the defect of the averaged longitudinal velocity component; $U$, $V$ and $W$ are the mean f\/low velocity component along $x$-, $y$- and $z$-axes, respectively; $\langle p_1 \rangle$ is the deviation from the hydrostatic pressure due to stratif\/ication $\rho_s(z)$; $g$ is the gravity acceleration; $\langle \rho_1 \rangle$ is the averaged density defect: $\rho_1=\rho-\rho_s$; $\rho_s=\rho_s(z)$ is the undisturbed f\/luid density assumed to be linear: $\rho_s(z)=\rho_0(1-az)$, $a>0$ is a constant; the prime indicates the pulsating components; $\langle \ \rangle$ indicates averaging.

In \cite{Hassid, Ch_NS2} the Reynolds stress tensor components $\langle u_i'u_j'\rangle$, the turbulent f\/lows $\langle u'_i\rho' \rangle$, and the density f\/luctuation variance $\langle \rho'^2 \rangle$ are def\/ined by the algebraic relations~\cite{Rodi}. Since the f\/low in the far turbulent wake is considered, these relations are simplif\/ied as follows
\begin{gather}
\langle u' v' \rangle=\frac{1-c_2}{c_1}\frac{e\langle v'^2\rangle}{\varepsilon}\frac{\partial U_d}{\partial y}=K_y\frac{\partial U_d}{\partial y},\label{equation9+}\\
\langle u'w'\rangle=\frac{(1-c_2)e\langle w'^2 \rangle-\displaystyle\frac{\mathstrut(1-c_3)(1-c_{2T})}{c_{1T}}\frac{g}{\rho_0}\frac{e^2}{\varepsilon}\langle w'\rho' \rangle}{c_1\varepsilon\left(1-\displaystyle\frac{\mathstrut 1-c_3}{c_1c_{1T}}\frac{g}{\rho_0}\frac{e^2}{\varepsilon^2}\frac{\partial\langle\rho\rangle}{\partial z} \right)}\frac{\partial U_d}{\partial z}=K_z\frac{\partial U_d}{\partial z},\label{equation10}\\
\langle v'^2 \rangle =\frac{2}{3}e\left(1-\frac{1-c_2}{c_1}\frac{P}{\varepsilon}-\frac{1-c_2}{c_1}\frac{G}{\varepsilon}\right),\label{equation11}\\
\langle w'^2 \rangle =\frac{2}{3}e\left(1-\frac{1-c_2}{c_1}\frac{P}{\varepsilon}+2\frac{1-c_2}{c_1}\frac{G}{\varepsilon}\right),\label{equation12}\\
\langle \rho'^2 \rangle =\frac{2}{c_T}\frac{e}{\varepsilon}\langle w'\rho' \rangle\frac{\partial \langle \rho \rangle}{\partial z},\label{equation13}\\
-\langle u'\rho' \rangle =\frac{1}{c_{1T}}\frac{e}{\varepsilon}\left( \langle u'w'\rangle \frac{\partial \langle \rho \rangle}{\partial z} +(1-c_{2T}) \langle w'\rho'\rangle \frac{\partial \langle U \rangle}{\partial z}\right),\label{equation14}\\
-\langle v'\rho' \rangle =\frac{1}{c_{1T}}\frac{e}{\varepsilon}\langle v'^2\rangle\frac{\partial \langle \rho \rangle}{\partial y}=K_{\rho y}\frac{\partial \langle \rho \rangle}{\partial y},\label{equation15}\\
-\langle w'\rho' \rangle =\frac{e}{c_{1T}\varepsilon}\left( \langle w'^2\rangle \frac{\partial \langle \rho \rangle}{\partial z} +(1-c_{2T})\frac{g}{\rho_0} \langle \rho'^2\rangle\right)\nonumber\\
\hphantom{-\langle w'\rho' \rangle}{}
=\frac{e\langle w'^2\rangle}{c_{1T}\varepsilon\left(1-2\displaystyle\frac{\mathstrut 1-c_{2T}}{c_{1T}c_{2T}}\frac{g}{\rho_0}\frac{e^2}{\varepsilon^2}\frac{\partial \langle\rho\rangle}{\partial z}\right)}\frac{\partial \langle \rho \rangle}{\partial z}=K_{\rho z}\frac{\partial \langle \rho \rangle}{\partial z},\label{equation17}\\
P=\left(\langle u'v'\rangle\frac{\partial U_d}{\partial y}+\langle u'w'\rangle\frac{\partial U_d}{\partial z}\right),\label{equation18}\\
G=-\frac{1}{\rho_0}\langle w'\rho' \rangle g.\label{equation19}
\end{gather}

Dif\/ferential transport equations \cite{Rodi} are used in \cite{Ch2,Ch1,Ch3} to determine the kinetic turbulent energy $e$, the kinetic energy dissipation rate $\varepsilon$, and the shear Reynolds stress $v'w'$:
\begin{gather}
U_0\frac{\partial e}{\partial x}+V\frac{\partial e}{\partial y}+W\frac{\partial e}{\partial z}=\frac{\partial}{\partial y}K_{ey}\frac{\partial e}{\partial y}+\frac{\partial}{\partial z}K_{ez}\frac{\partial e}{\partial z}+P+G-\varepsilon,\label{equation20}\\
U_0\frac{\partial \varepsilon}{\partial x}+V\frac{\partial \varepsilon}{\partial y}+W\frac{\partial \varepsilon}{\partial z}=\frac{\partial}{\partial y}K_{\varepsilon y}\frac{\partial \varepsilon}{\partial y}+\frac{\partial}{\partial z}K_{\varepsilon z}\frac{\partial \varepsilon}{\partial z}+c_{\varepsilon 1}\frac{\varepsilon}{e}(P+G)-c_{\varepsilon 2}\frac{\varepsilon^2}{e},\label{equation21}\\
U_0\frac{\partial \langle v'w' \rangle}{\partial x}+V\frac{\partial \langle v'w' \rangle}{\partial y}+W\frac{\partial \langle v'w' \rangle}{\partial z}=\frac{\partial}{\partial y}K_{ey}\frac{\partial\langle v'w' \rangle}{\partial y}+\frac{\partial}{\partial z}K_{ez}\frac{\partial\langle v'w' \rangle}{\partial z}\nonumber\\
\qquad{} +(c_2-1)\left( \langle v'^2 \rangle\frac{\partial W}{\partial y}+\langle w'^2 \rangle\frac{\partial V}{\partial z} \right)+(1-c_3)\frac{g}{\rho_0}\langle v'\rho' \rangle - c_1\frac{\varepsilon}{e}\langle v'w' \rangle.\label{equation22}
\end{gather}
The turbulent viscosity coef\/f\/icients in these equations are $K_{ey}=K_y$, $K_{ez}=K_z$, $K_{\varepsilon y}=K_{ey}/\delta$, $K_{\varepsilon z}=K_{ez}/\delta$. The model \eqref{equation1}--\eqref{equation4} is an analogue of the equations \eqref{equation5}--\eqref{equation22} (for the dif\/fusion approximation in a homogeneous f\/luid $V=0$, $W=0$, and $g=0$).
 In what follows, we assume that the free stream velocity $U_0$ equals unity.

\section{Similarity solution}\label{section3}

The inf\/initesimal symmetry group \cite{Olver, LV} of the model \eqref{equation1}--\eqref{equation4} is spanned by the eight vector f\/ields
\begin{gather}
X_1=\frac{\partial}{\partial x},\!\qquad X_2=\frac{\partial}{\partial y},\!\qquad X_3=\frac{\partial}{\partial z},\!\qquad X_4=\frac{\partial}{\partial \langle\rho_1\rangle},\!\qquad X_5=-z\frac{\partial}{\partial y}+y\frac{\partial}{\partial z}+y\frac{\partial}{\partial \langle\rho_1\rangle}, \nonumber\\
X_6=y\frac{\partial}{\partial y}+z\frac{\partial}{\partial z}+2e\frac{\partial}{\partial e}+2\varepsilon\frac{\partial}{\partial \varepsilon}+\langle\rho_1\rangle\frac{\partial}{\partial \langle\rho_1\rangle}+2\langle\rho'^2\rangle\frac{\partial}{\partial \langle\rho'^2\rangle},\nonumber\\
X_7=x\frac{\partial}{\partial x}-2e\frac{\partial}{\partial e}-3\varepsilon\frac{\partial}{\partial \varepsilon},\qquad X_8=(\langle\rho_1\rangle-z)\frac{\partial}{\partial \langle\rho_1\rangle}+2\langle\rho'^2\rangle\frac{\partial}{\partial \langle\rho'^2\rangle}.\nonumber
\end{gather}
Available experimental data \cite{Hinze,LP} and numerical calculations \cite{Hassid,Pope,Wilcox} show that the f\/low in the far turbulent wake can be considered to be close to a self-similar f\/low. We therefore consider the linear combination of scaling vector f\/ields $X_6$ and $X_7$
\[
Z=x\frac{\partial}{\partial x}+\alpha y\frac{\partial}{\partial y}+\alpha z\frac{\partial}{\partial z}+2(\alpha-1)e\frac{\partial}{\partial e}+(2\alpha-3)\varepsilon\frac{\partial}{\partial \varepsilon}+\alpha\langle\rho_1\rangle\frac{\partial}{\partial\langle\rho_1\rangle}+2\alpha\langle\rho'^2\rangle\frac{\partial}{\partial \langle\rho'^2\rangle}.
\]
The similarity variables associated with the inf\/initesimal generator $Z$ are
\begin{gather*}
\xi=\frac{y}{x^{\alpha}},\quad \eta=\frac{z}{x^{\alpha}},\qquad e=x^{2\alpha-2}E(\xi,\eta),\qquad \varepsilon=x^{2\alpha-3}G(\xi,\eta),\nonumber\\
\langle\rho_1\rangle=x^\alpha H(\xi,\eta),\qquad \langle\rho'^2\rangle=x^{2\alpha} R(\xi,\eta),\nonumber
\end{gather*}
where $E$, $G$, $H$, and $R$ are arbitrary functions. According to physical considerations (the inf\/luence of gravity is neglected in this problem), the functions $E$ and $G$ must be of the form
\begin{gather}
E(\xi,\eta)=E\Big(\sqrt{\xi^2+\eta^2}\Big),\qquad G(\xi,\eta)=G\Big(\sqrt{\xi^2+\eta^2}\Big).\label{equation23}
\end{gather}

We obtain the reduced system by introducing similarity variables. Using \eqref{equation23} and changing to polar coordinates $\xi=r\cos\phi$, $\eta=r\sin\phi$ the reduced system becomes
\begin{gather}
C_e \frac{E}{G}\left(EE''+2E'^2-\frac{E}{G}E'G'+\frac{E}{r}E'\right)+\alpha rE'+2(1-\alpha)E-G=0,\label{equation24}\\
C_{\varepsilon} \frac{E}{G}\left(EG''-\frac{E}{G}G'^2+2E'G'+\frac{E}{r}G'\right)+\alpha rG'+(3-2\alpha)G-{C_{\varepsilon}}_2\frac{G^2}{E}=0,\label{equation25}\\
C_\rho \frac{E^2}{G}\left(H_{rr}+\frac{1}{r^2}H_{\phi\phi}\right)+\left(C_\rho \frac{E}{G}\left(2E'-\frac{E}{G}G'+\frac{E}{r} \right)+\alpha r\right)H_r-\alpha H\nonumber\\
\qquad{}+C_\rho \frac{E}{G}\left(\frac{E}{G}G'-2E'\right)\sin\phi=0,\label{equation26}\\
{{C_1}_\rho} \frac{E^2}{G}\left(R_{rr}+\frac{1}{r^2}R_{\phi\phi}\right)+\left({{C_1}_\rho} \frac{E}{G}\left( \frac{E}{r}+2E'-\frac{E}{G}G'\right)+\alpha r\right)R_r-\left(C_T \frac{E}{G}+2\alpha\right)R\nonumber\\
\qquad{}+2C_\rho \frac{E^2}{G}\left(H_{r}^2+\frac{1}{r^2}H_{\phi}^2\right)+2C_\rho \frac{E^2}{G}-4C_\rho \frac{E^2}{G}\left(H_r\sin\phi+H_\phi\frac{\cos\phi}{r}\right)=0,\label{equation27}
\end{gather}
where $E=E(r)$, $G=G(r)$, $H=H(r,\phi)$, and $R=R(r,\phi)$. Here, and throughout, subscripts denote derivatives, so $H_r=\partial H/ \partial r$, etc.

Lie's classical method do not provide solution of the reduced system agreed with experimental data. We therefore use the BDEs method.

\section{BDEs method}\label{section4}

The concept of BDEs of a system of partial dif\/ferential equations (PDEs) was introduced in~\mbox{\cite{BDE,BDE_Paper}}. Consider a scalar PDE
\begin{gather}
\Omega\big({\bf{x}},u,u^1,u^2,\dots\big)=0,\label{PDE}
\end{gather}
where ${\bf{x}}=(x_1,x_2,\dots,x_n)$ denotes $n$ independent variables, $u$ denotes the dependent variable, and
\[
u_k=\frac{\partial u^k}{\partial x_{i_1}\partial x_{i_2}\cdots\partial x_{i_k}}
\]
denotes the set of coordinates corresponding to all $k$-th order partial derivatives of $u$ with respect to ${\bf{x}}$. In the BDEs method, an extension of the classical symmetry determining relations is made by incorporating an additional factor $B({\bf{x}},u,u^1,u^2,\dots)$. For a scalar PDE (\ref{PDE}), BDE is
\begin{gather}
h\frac{\partial \Omega}{\partial u}+\sum_{|\alpha|\ge 1}D^\alpha (h) \frac{\partial \Omega}{\partial u_\alpha}+Bh|_{\Omega=0}=0, \label{Gen_BDE}
\end{gather}
where $h$ is a function of ${\bf{x}},u,u^1,u^2,\dots$; $D^\alpha=D^{\alpha_1}_{x_1}\cdots D^{\alpha_n}_{x_n}$, $D_{x_i}$ is the total $x_i$ derivative; $\alpha$ is a~multi-index.  Equality (\ref{Gen_BDE}) must hold for all solutions of~(\ref{PDE}).

Now we use the BDEs method to reduce \eqref{equation26} and \eqref{equation27} to some ODEs. Consider more general equation than \eqref{equation26}
\begin{gather}
H_{\phi\phi}+r^2H_{rr}+A(r)H_r+B(r)H+C(r)\sin\phi=0,\label{equation30}
\end{gather}
where $A(r)$, $B(r)$, and $C(r)$ are arbitrary functions. BDE corresponding to \eqref{equation30} is
\begin{gather}
D_\phi^2 h+r^2D_r^2h+b_1(r,\phi)D_rh+b_2(r,\phi)h=0.\label{equation31}
\end{gather}
Here, and throughout $D_\phi$, $D_r$ are the operators of total dif\/ferentiation with respect to~$\phi$ and $r$. The function~$h$ may depend on~$r$, $\phi$, $H$ and derivatives of $H$. The functions $b_1(r,\phi)$ and $b_2(r,\phi)$ are to be determined together with the function $h$. Note that if we let in~\eqref{equation31}
\[
b_1(r,\phi)=A(r), \qquad b_2(r,\phi)=B(r),
\]
we obtain the classical determining equation~\cite{Olver, LV}.

For simplicity, we assume that a solution of \eqref{equation31} is independent of $r$ and partial derivatives of~$H$ with respect to~$r$
\begin{gather}
h=H_{\phi\phi}+h_1\left(\phi,H,H_\phi\right).\label{equation32}
\end{gather}
Substituting \eqref{equation32} into \eqref{equation31} gives a polynomial equation for derivatives of the fourth order. This polynomial must identically vanish. We can express the derivatives $H_{rr\phi\phi}$, $H_{\phi\phi\phi\phi}$, $H_{r\phi\phi}$, $H_{\phi\phi\phi}$, and $H_{\phi\phi}$ using \eqref{equation30}. The coef\/f\/icient of $H_{rrr}$ implies $b_1(r,\phi)=A(r)$.

As a result, the left side of \eqref{equation31} is a polynomial in $H_{rr}$ and $H_{r\phi}$.  Collecting similar terms we obtain
\begin{gather}
2\left(A(r)H_r+B(r)H+C(r)\sin\phi \right)h_{1_{H_\phi H_\phi}}-2H_\phi h_{1_{H H_\phi}}-2h_{1_{\phi H_\phi}}+B(r)-b_2(r,\phi)=0,\nonumber\\
h_{1_{H_\phi H_\phi}}=0, \qquad h_{1_{H H_\phi}}=0.\nonumber
\end{gather}
Hence
\[
h_1\left(\phi,H,H_\phi\right)=h_2(\phi)H_\phi+h_3(\phi,H), \qquad b_2(r,\phi)=B(r)-2h_2'(\phi).
\]

Substituting the functions $b_1$, $b_2$, and $h_1$ into the left side of \eqref{equation31} we obtain the polynomial with respect to $H_{r}$ and $H_\phi$. This polynomial must identically vanish. Collecting similar terms we have
\begin{gather}
(B(r)H+C(r)\sin\phi)h_{3_H}-h_{3_{\phi\phi}}+(2h_2'-B(r))h_3+C(r)(h_2\cos\phi-\sin\phi)=0,\nonumber\\
h_{3_{HH}}=0, \qquad 2h_{3_{\phi H}}+h_2''-2h_2'h_2=0.\nonumber
\end{gather}
Thus,{\samepage
\begin{gather}
h_3(\phi,H)=\left(\frac{1}{2}h_2^2-\frac{1}{2}h_2'+h_4\right)H,\qquad
h_2'-h_2^2-2\cot\phi h_2+2(1-h_4)=0,\label{equation33}
\end{gather}
where $h_4$ is an arbitrary constant.}

Clearly, that the Riccati equation \eqref{equation33} has the partial solution
\[
h_2=\tan\phi
\]
for $h_4=1/2$.

Thus we f\/ind the solution of \eqref{equation31}
\[
h=H_{\phi\phi}+H_\phi\tan\phi.
\]
The corresponding dif\/ferential constraint $h=0$ has the general solution
\begin{gather}
H=H_1(r)\sin\phi+H_2(r),\label{equation34}
\end{gather}
where $H_1$ and $H_2$ are arbitrary functions.

Next we use \eqref{equation34} and consider the equation \eqref{equation27} in more general form
\begin{gather}
R_{\phi\phi}+r^2R_{rr}+K(r)R_r+L(r)R+M(r)\sin^2\phi+N(r)\sin\phi+P(r)=0,\label{equation35}
\end{gather}
where $K(r)$, $L(r)$, $M(r)$, $N(r)$, and $P(r)$ are arbitrary functions. The BDEs method applied to the equation \eqref{equation35} gives rise to the following results:
\begin{gather*}
N(r)=0,\qquad
b_1(r,\phi)=K(r), \qquad b_2(r,\phi)=L(r)-8\sin^{-2}2\phi,\nonumber\\
h=R_{\phi\phi}-2R_{\phi}\cot 2\phi.%\label{equation36}
\end{gather*}
Integrating dif\/ferential constraint $h=0$ corresponding to the BDE solution \eqref{equation34}, we f\/ind
\begin{gather}
R=R_1(r)\sin^2\phi+R_2(r),\label{equation37}
\end{gather}
where $R_1(r)$ and $R_2(r)$ are arbitrary functions.
Using \eqref{equation34} and \eqref{equation37} we obtain the following corollary in terms of variables $x$, $y$, and $z$.

\begin{corollary}
The following expressions for the unknown functions
\begin{gather}
e=x^{2\alpha-2}E\left(\frac{\sqrt{y^2+z^2}}{x^\alpha}\right),\qquad \varepsilon=x^{2\alpha-3}G\left(\frac{\sqrt{y^2+z^2}}{x^\alpha}\right),\label{equation38}\\
\langle\rho_1\rangle=zH\left(\frac{\sqrt{y^2+z^2}}{x^\alpha}\right),\qquad \langle\rho'^2\rangle=z^2R_1\left(\frac{\sqrt{y^2+z^2}}{x^\alpha}\right) + x^{2\alpha}R_2\left(\frac{\sqrt{y^2+z^2}}{x^\alpha}\right)\label{equation39}
\end{gather}
allow us to reduce the model \eqref{equation1}--\eqref{equation4} to a system of ODEs.
\end{corollary}

Indeed, substituting \eqref{equation38} and \eqref{equation39} into \eqref{equation1}--\eqref{equation4} we obtain
\begin{gather}
\frac{E^2E''}{G}-\frac{E^2E'}{G}\left(\frac{G'}{G}-2\frac{E'}{E}-\frac{1}{\tau}\right)-\frac{1}{C_e}\left(2(\alpha-1)E+G-\alpha\tau E'\right)=0,\label{equation40}\\
\frac{E^2G''}{G}-\frac{E^2G'}{G}\left(\frac{G'}{G}-2\frac{E'}{E}-\frac{1}{\tau}\right) -\frac{1}{C_{\epsilon}}\left((2\alpha-3)G+{C_{\epsilon}}_2\frac{G^2}{E}-\alpha\tau G'\right)=0,\label{equation41}\\
\frac{E^2H''}{G}-\frac{E^2H'}{G}\left(\frac{G'}{G}-2\frac{E'}{E}-\frac{3}{\tau}-\frac{\alpha}{C_\rho}\frac{\tau G}{E^2}\right)-\frac{E^2}{\tau G}\left(\frac{G'}{G}-2\frac{E'}{E}\right)(H-1)=0,\label{equation42}\\
\frac{E^2R_1''}{G}-\frac{E^2R_1'}{G}\left(\frac{G'}{G}-2\frac{E'}{E}-\frac{5}{\tau}-\frac{\alpha}{{C_1}_\rho}\frac{\tau G}{E^2}\right)-\frac{E^2R_1}{\tau G}\left(2\frac{G'}{G}-4\frac{E'}{E}+\frac{C_T}{{C_1}_\rho}\frac{\tau G}{E^3}\right)\nonumber\\
\qquad{}+2\frac{C_\rho}{{C_1}_\rho}\frac{E^2H'}{G}\left(2\frac{(H-1)}{\tau}+H'\right)=0,\label{equation43}\\
\frac{E^2R_2''}{G}-\frac{E^2R_2'}{G}\left(\frac{G'}{G}-2\frac{E'}{E}-\frac{1}{\tau}-\frac{\alpha}{{C_1}_\rho}\frac{\tau G}{E^2}\right)-\frac{R_2}{{C_1}_\rho}\left(C_T\frac{G}{E}+2\alpha\right)+2\frac{E^2R_1}{G}\nonumber\\
\qquad{}+2\frac{C_\rho}{{C_1}_\rho}(H-1)^2\frac{E^2}{G}=0,\label{equation44}
\end{gather}
where $\tau=\sqrt{\xi^2+\eta^2}$.

\section{Calculation results}\label{section5}

The system of ODEs \eqref{equation40}--\eqref{equation44}  has to satisfy the conditions
\begin{gather}
E'=G'=H_1'=R_1'=R_2'=0,\qquad \tau=0,\label{equation45}\\
E=G=H_1=R_1=R_2=0,\qquad \tau\to\infty.\label{equation46}
\end{gather}
Conditions \eqref{equation45} take into account f\/low symmetry with respect to the OX axis. The boundary conditions \eqref{equation46} imply that all functions take zero values outside the turbulent wake.

\begin{figure}[t]\small \centering
\begin{minipage}[h]{0.40\linewidth}
\center{\includegraphics[width=1\linewidth]{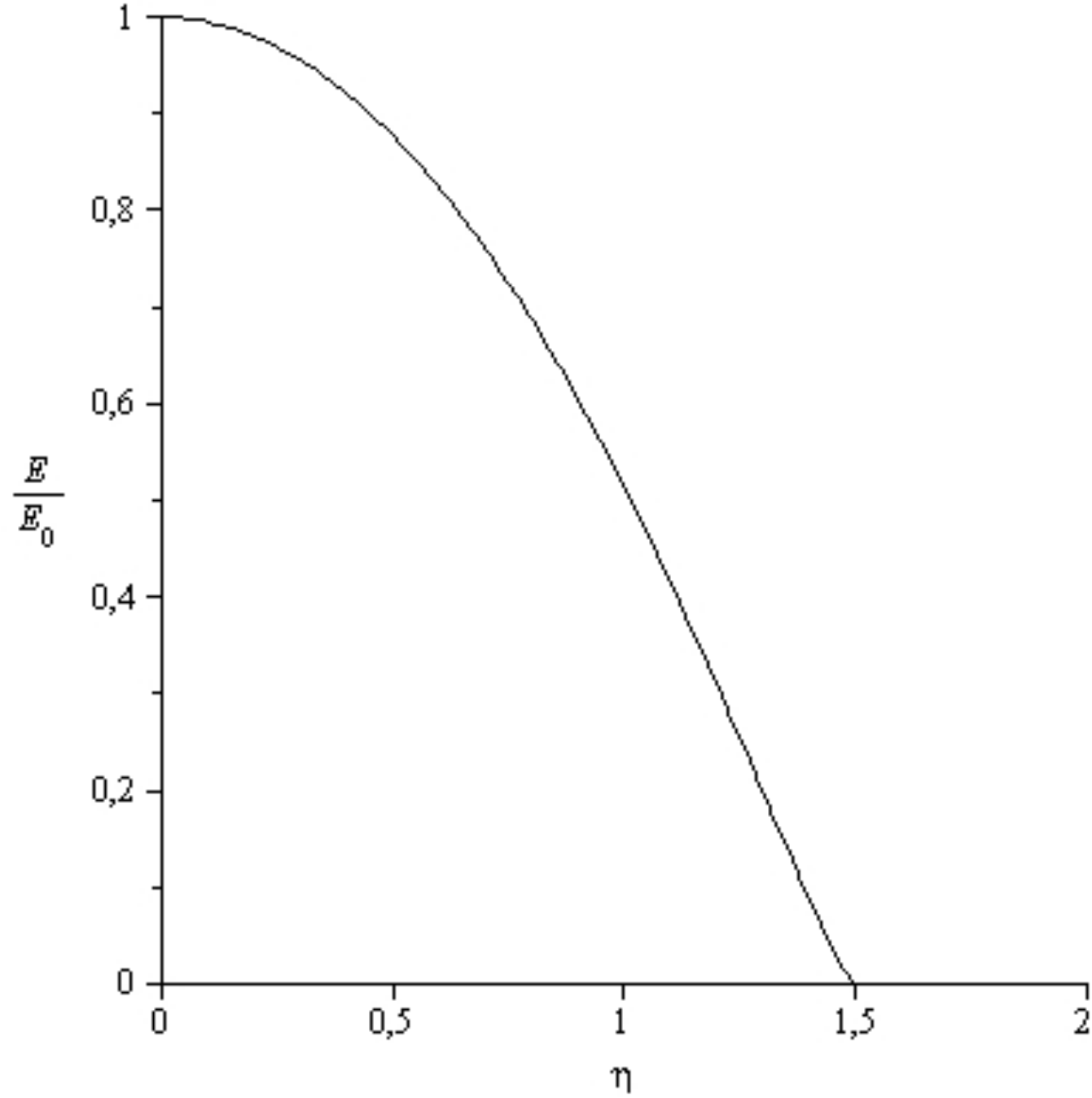}}\\ (a) \\
\end{minipage}
\qquad\quad
\begin{minipage}[h]{0.40\linewidth}
\center{\includegraphics[width=1\linewidth]{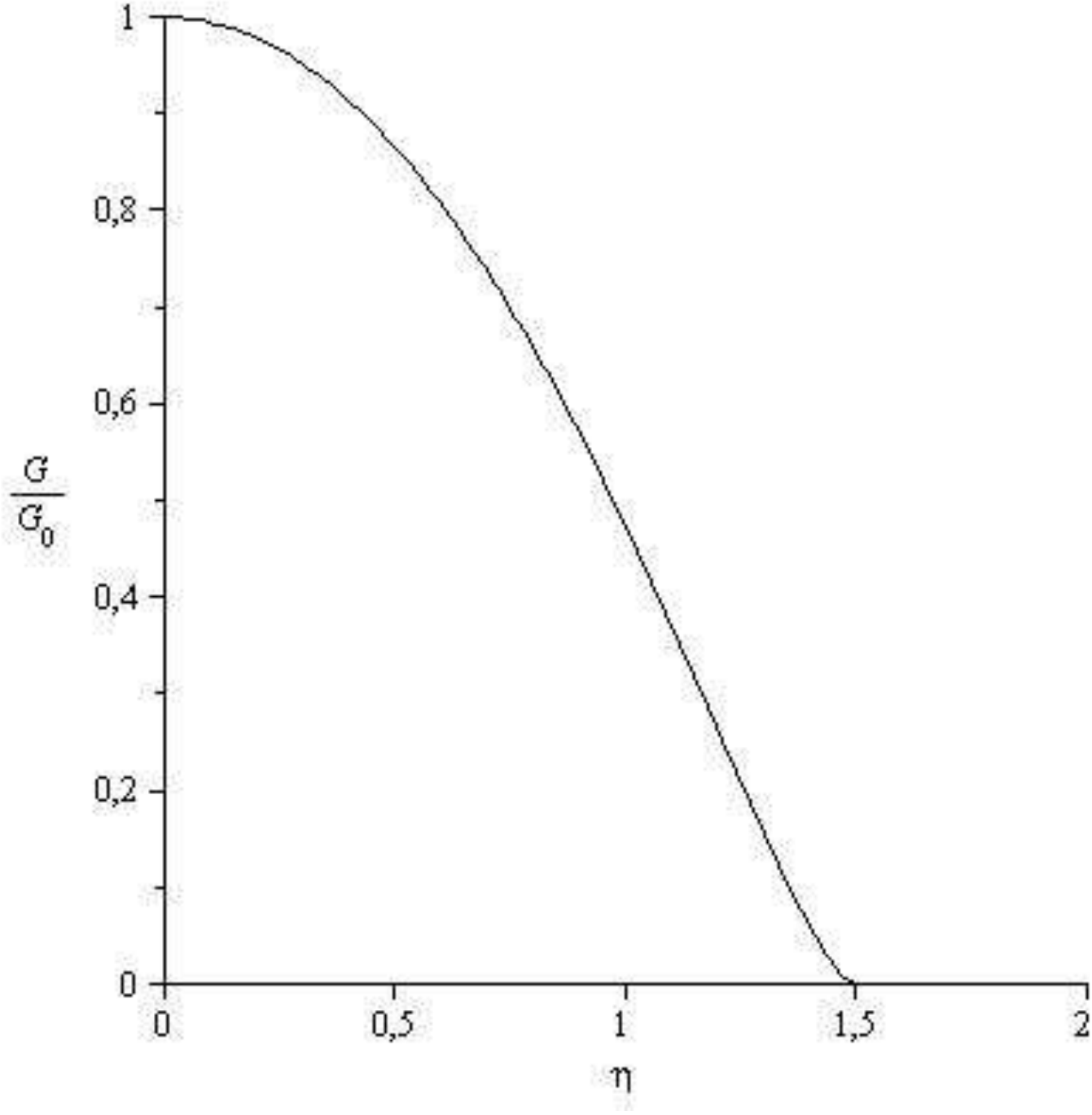}}\\ (b) \\
\end{minipage}
\medskip

\begin{minipage}[h]{0.40\linewidth}
\center{\includegraphics[width=1\linewidth]{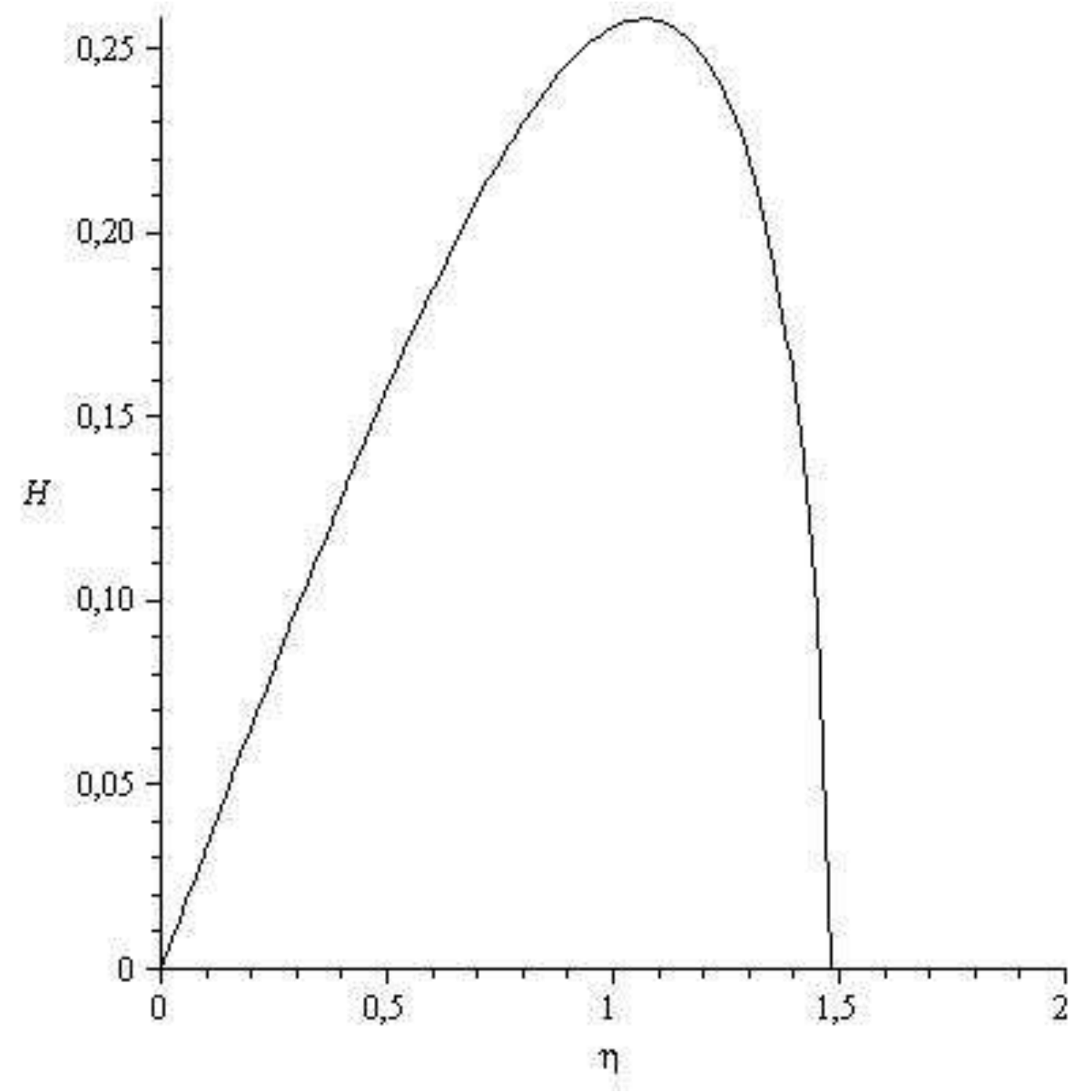}}\\ (c) \\
\end{minipage}
\qquad\quad
\begin{minipage}[h]{0.40\linewidth}
\center{\includegraphics[width=1\linewidth]{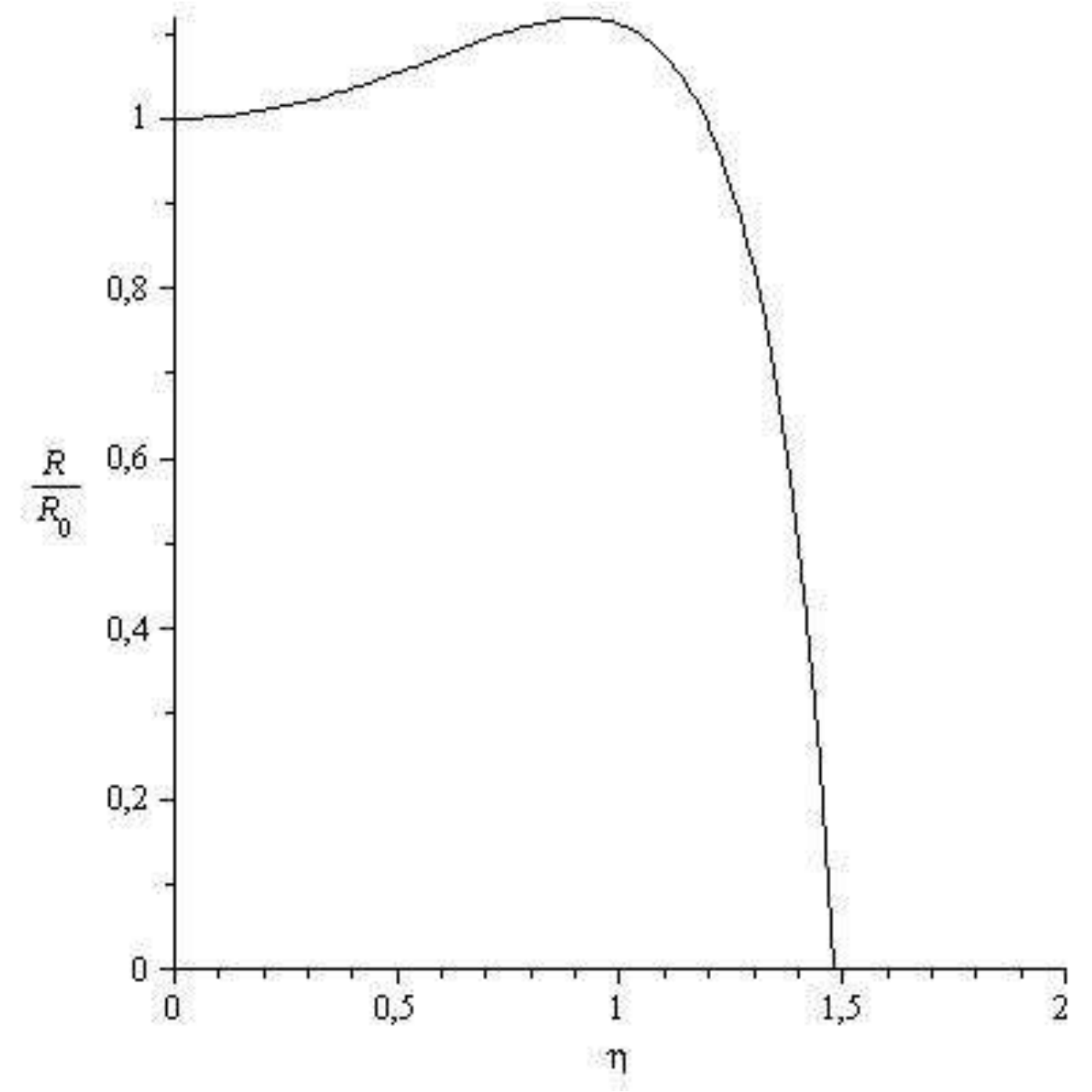}}\\ (d) \\
\end{minipage}
\caption{Calculated prof\/iles as $\xi=0$: (a) normed prof\/ile of $E$, (b) normed prof\/ile of $G$, (c) prof\/ile of~$H$, (d) normed prof\/ile of~$R$.}\label{Fig1}
\end{figure}

The system of ODEs \eqref{equation40}--\eqref{equation44} satisfying boundary conditions~\eqref{equation45},~\eqref{equation46} was solved numerically. Additional dif\/f\/iculties are caused by the fact that the coef\/f\/icients of ODEs have singula\-ri\-ties. The problem was solved by the modif\/ied shooting method and asymptotic expansion of the solution in the vicinity of the singular point~\cite{Cazalbou,Shanko}
\begin{gather}
E=c_1(\tau-a)^{10/7}+o\big(|\tau-a|^{10/7}\big),\qquad G=-\frac{30C_e c_1^2}{7a}(\tau-a)^{13/7}+o\big(|\tau-a|^{13/7}\big),\nonumber\\
H=\frac{7C_\rho}{a(7C_\rho-10C_e)}(\tau-a)+o(|\tau-a|),\nonumber\\
R_1=\frac{49C_\rho^2(7(2a+1)C_\rho-20aC_e)}{2a^2(7C_\rho-10C_e)^2(5C_e-7{C_\rho}_1)}(\tau-a)^2+o\big(|\tau-a|^2\big),\nonumber\\
R_2=\frac{7C_\rho}{2(5C_e-7{C_1}_\rho)}(\tau-a)^2+o\big(|\tau-a|^2\big).\nonumber
\end{gather}

\begin{figure}[t]\small \centering
\begin{minipage}[h]{0.45\linewidth}
\center{\includegraphics[height=63mm]{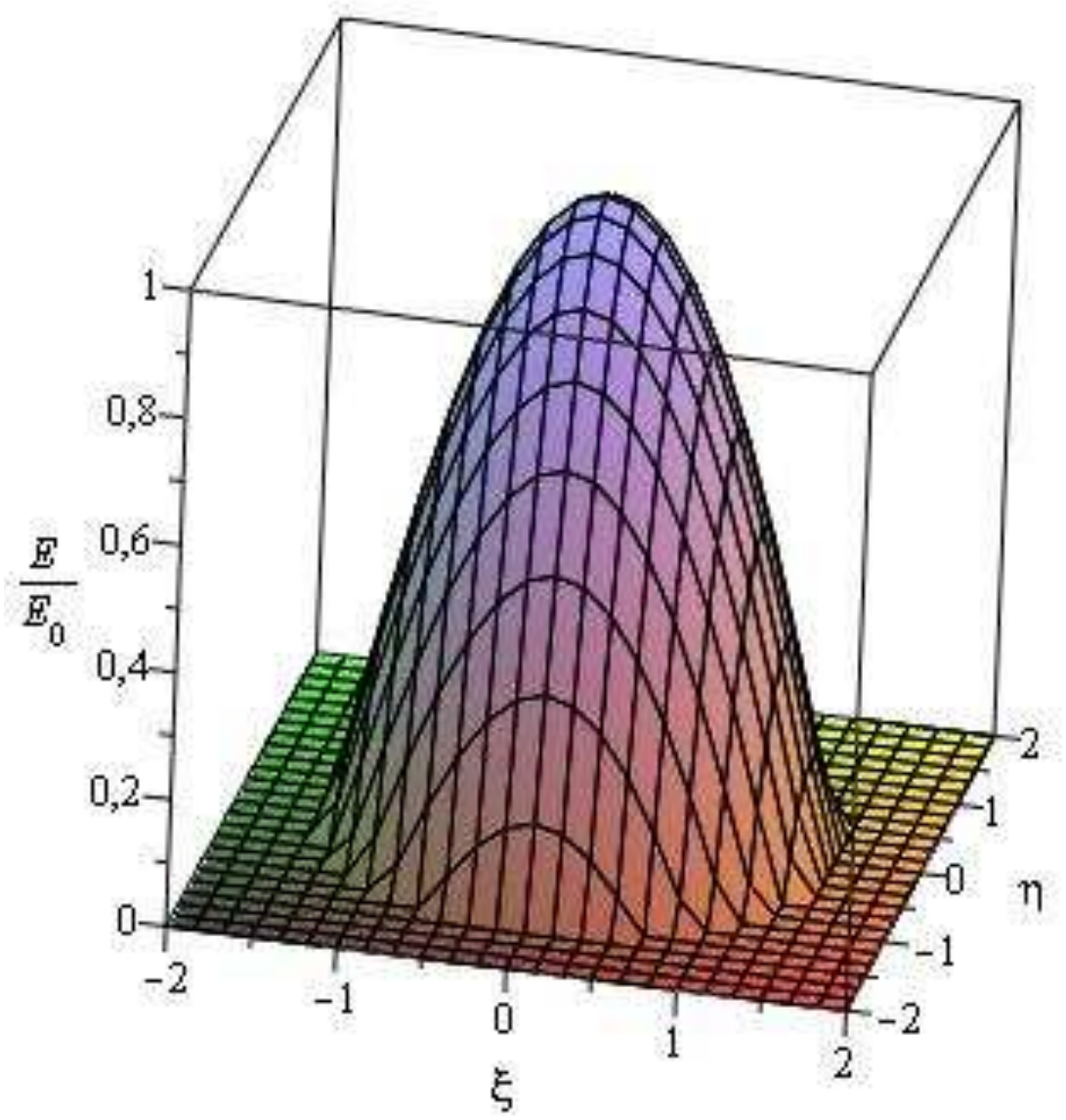}} \\
(a) \\
\end{minipage}
\qquad\quad
\begin{minipage}[h]{0.45\linewidth}
\center{\includegraphics[height=63mm]{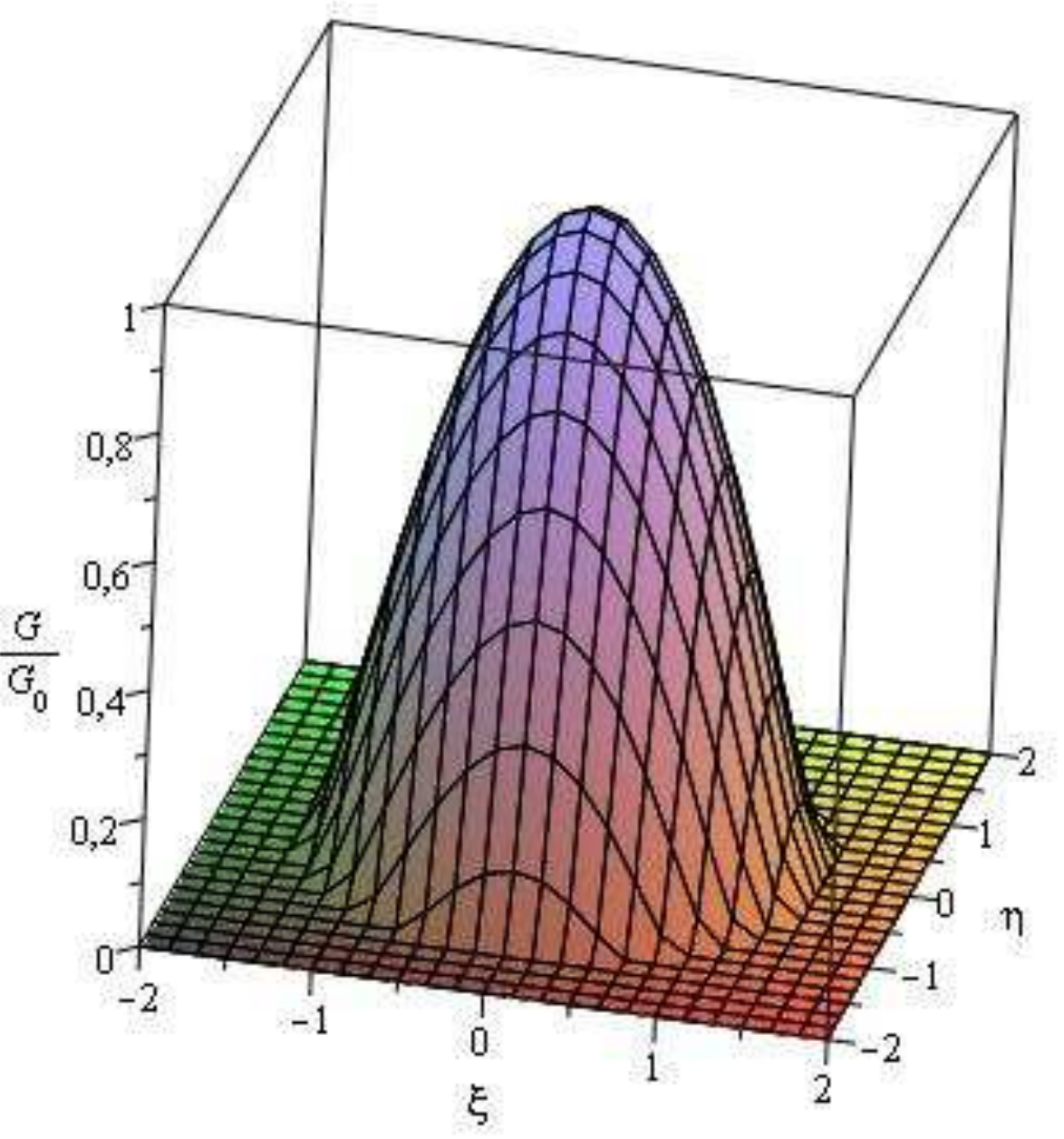}}\\ (b) \\
\end{minipage}
\medskip

\begin{minipage}[h]{0.45\linewidth}
\center{\includegraphics[height=63mm]{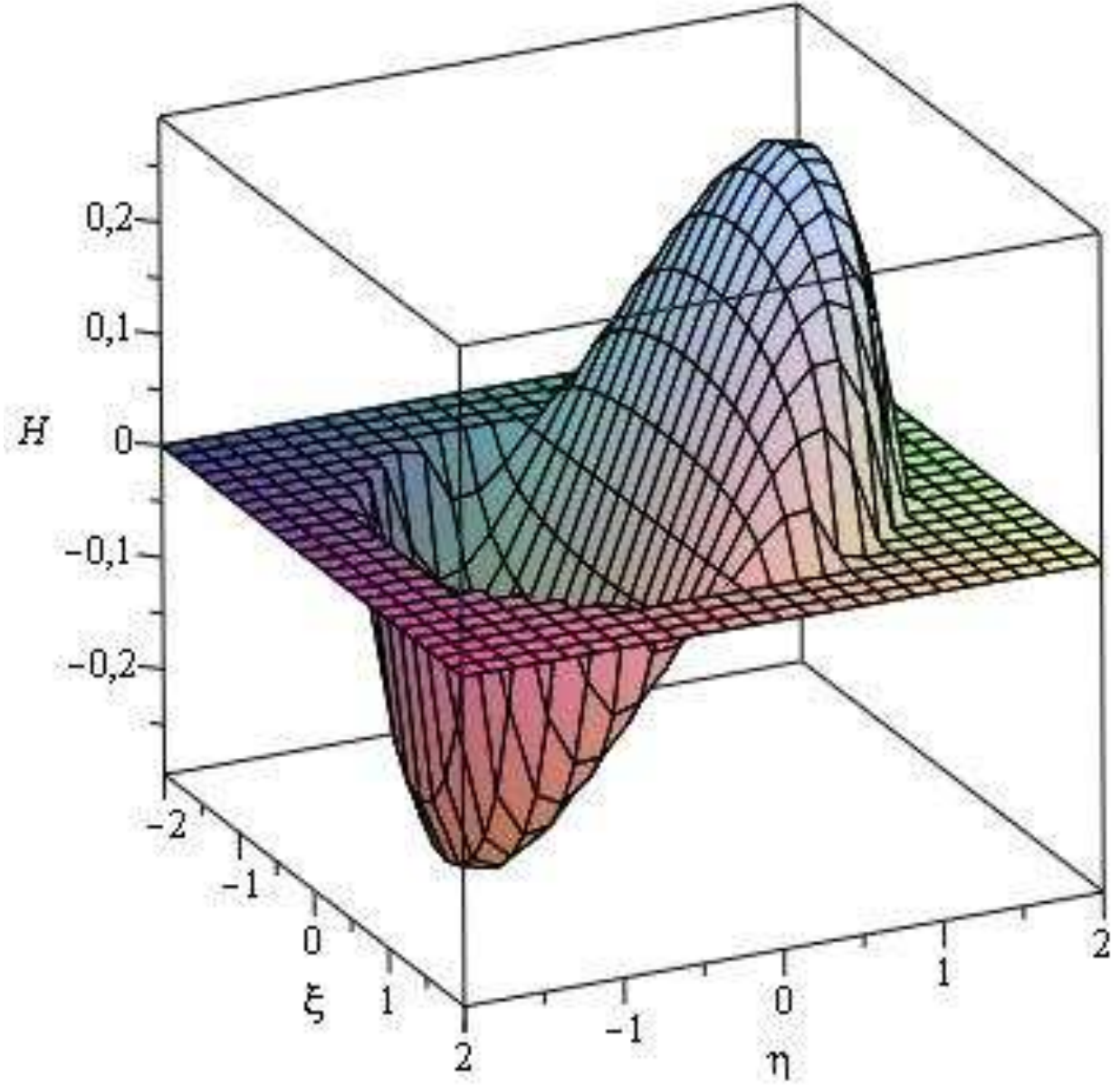}}\\ (c) \\
\end{minipage}
\qquad\quad
\begin{minipage}[h]{0.45\linewidth}
\center{\includegraphics[height=63mm]{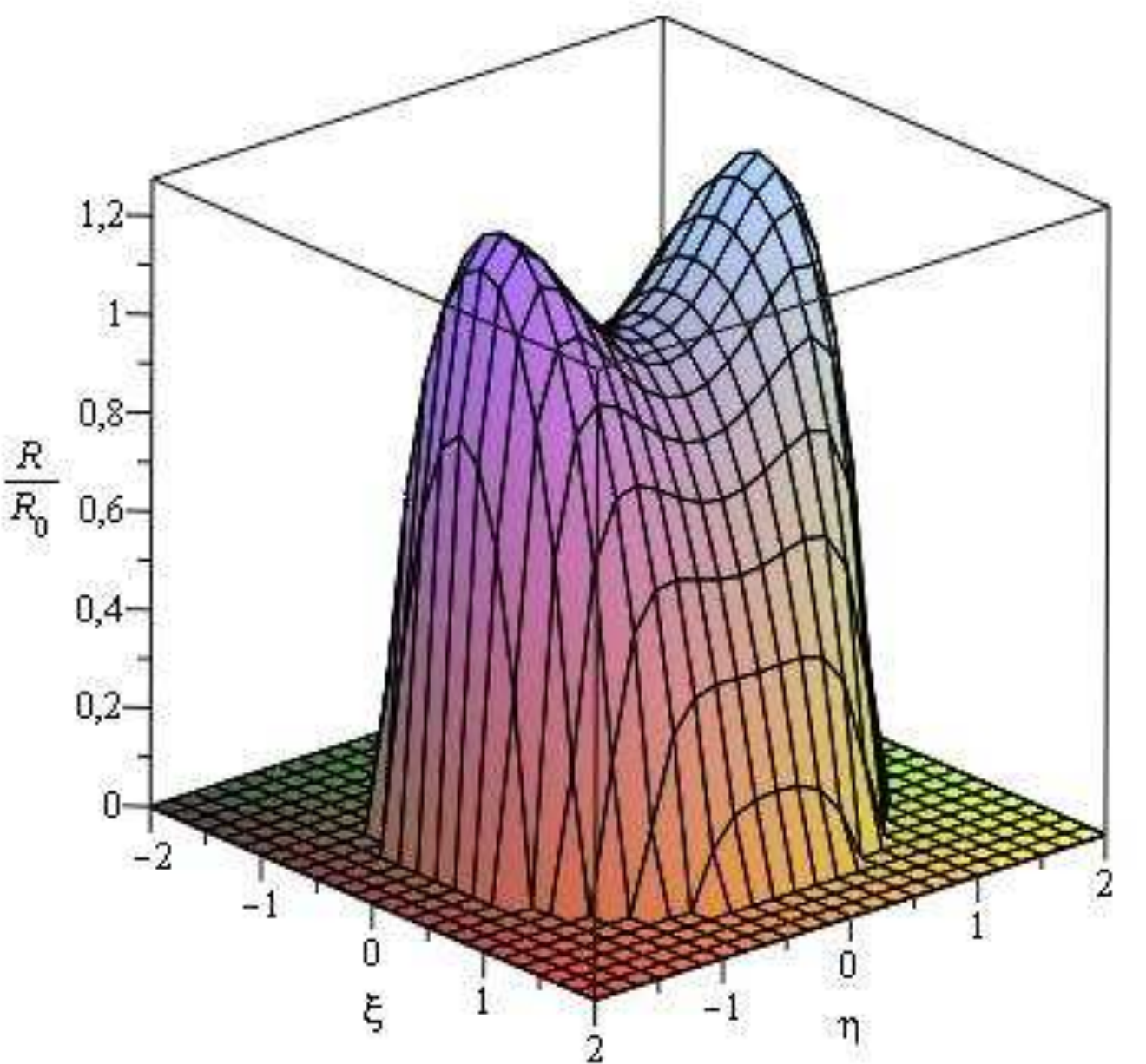}}\\ (d) \\
\end{minipage}
\caption{Calculated functions: (a) the function $E/E_0$, (b) the function $G/G_0$, (c) the function~$H$, (d) the function $R/R_0$.}\label{Fig2}
\end{figure}

The value of $\alpha$ is taken to be $0.23$ in accordance with experimental data~\cite{Hassid, LP}. The results for the problem solution are illustrated in Figs.~\ref{Fig1} and~\ref{Fig2}. Fig.~\ref{Fig1} shows the prof\/iles of the functions $E/E_0$, $G/G_0$, $H$, and $R/R_0$ as $\xi=0$, where subscript $0$ denotes the axial value. The functions $E/E_0$, $G/G_0$, $H$, and $R/R_0$ are plotted in Fig.~\ref{Fig2}. The functions $E/E_0$ and $G/G_0$ are bell-shaped and determine shapes of the normalized turbulent kinetic energy and the normalized kinetic energy dissipation rate respectively.  Similarly, $H$ and $R/R_0$ determine shapes of the average density defect and the normalized density f\/luctuation varience respectively.

The function $H(0,\eta)$ characterizing the degree of f\/luid mixing in the turbulent wake is given in Fig.~\ref{Fig1}c. The maximum value of this function equals $0.258$. This is in consistent with the present notions of incomplete f\/luid mixing in the wakes~\cite{TurbMix,Voropaeva}.

\begin{figure}[t]\centering
 \includegraphics[width=0.45\linewidth]{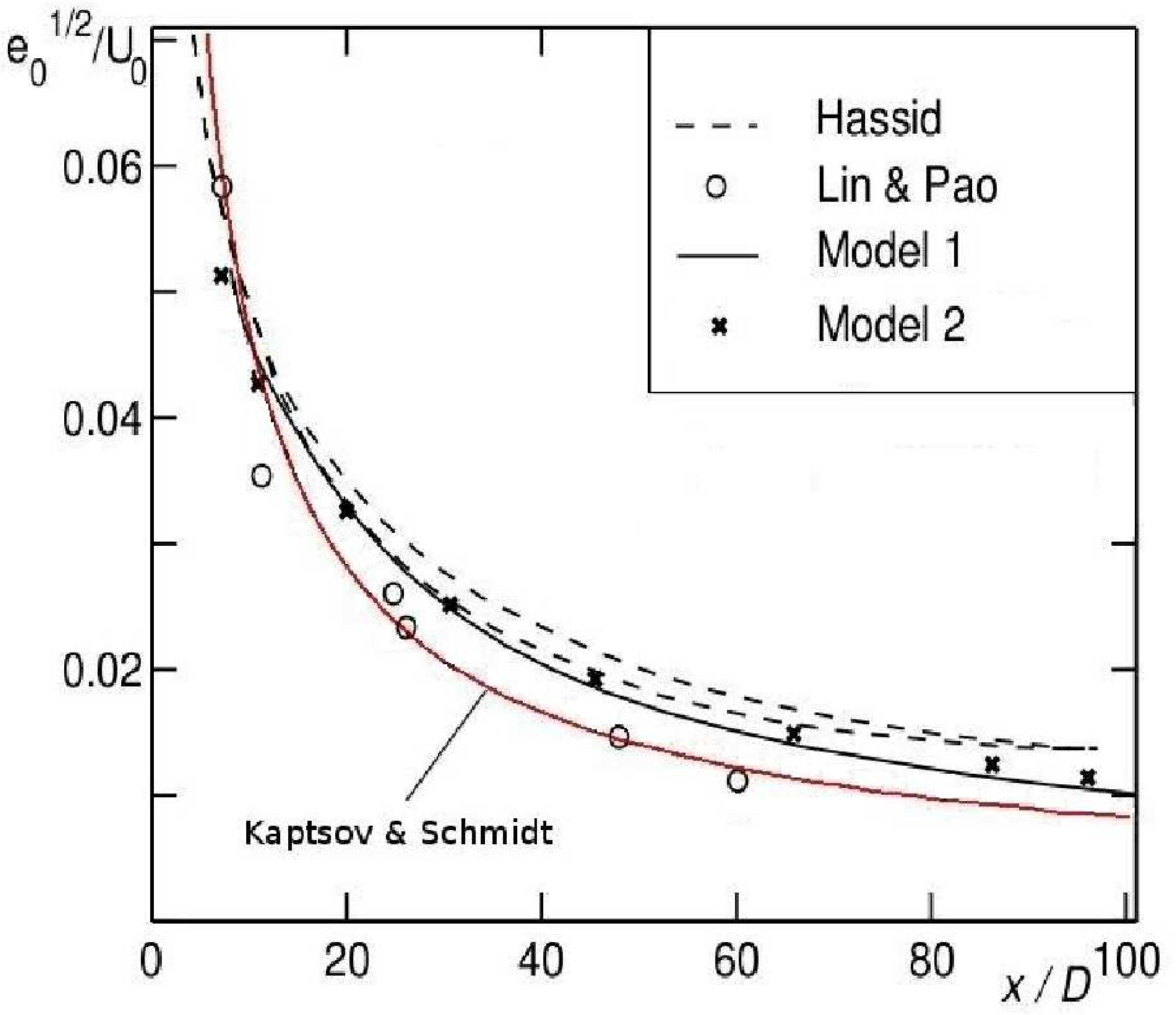}

\caption{Axial values of the turbulent energy.}\label{Fig3}
\end{figure}

In Fig.~\ref{Fig3} adapted from \cite{Ch1} the normalized values of the turbulent energy along the wake axis $e_0^{1/2}/U_0=e(x,0,0)/U_0$ are compared with experimental data \cite{LP}, computational results~\cite{Hassid} and results of numerical simulation based on two semi-empirical turbulence models (Model1 and Model2 in \cite{Ch2,Ch1}). The coordinate~$x$ is normalized by the body diameter $D$. The results obtained here are in close agreement with Lin and Pao's experimental data.

\subsection*{Acknowledgements}
The authors are grateful to Professor G.G.~Chernykh for many helpful and stimulating  discussions. The authors would like to thank unknown referees for valuable comments which corrected and improved the f\/irst version of this paper. This work was supported by the Russian Foundation for Basic Research (project no. 10-01-00435) and programme `Leading scientif\/ic schools' (grant no.~NSh-544.2012.1).

\pdfbookmark[1]{References}{ref}

\LastPageEnding

\end{document}